\documentclass[letterpaper,twocolumn,10pt]{article}
\usepackage{usenix2019_v3}

\usepackage{tikz}
\usepackage{amsmath}

\usepackage{filecontents}

\usepackage{amsmath, amssymb, amsfonts}
\usepackage{nicefrac}
\usepackage{balance}
\usepackage{algorithm}
\usepackage{algpseudocode}

\usepackage{seqsplit}
\makeatletter

\usepackage{booktabs}
\usepackage{tabularx}
\usepackage{array}
\usepackage{amssymb}

\usepackage{graphicx}
\makeatother
\AtBeginDocument{%
  \setlength{\abovedisplayskip}{3pt plus 1pt minus 1pt}%
  \setlength{\belowdisplayskip}{3pt plus 1pt minus 1pt}%
  \setlength{\abovedisplayshortskip}{1pt plus 1pt}%
  \setlength{\belowdisplayshortskip}{1pt plus 1pt}%
}
\makeatletter
\def\thm@space@setup{%
  \thm@preskip=2pt plus 1pt minus 1pt
  \thm@postskip=2pt plus 1pt minus 1pt}
\makeatother
\makeatletter
\@ifpackageloaded{enumitem}{\setlist{topsep=2pt,itemsep=2pt,parsep=0pt}}{}
\makeatother

\newcommand{\sys}{TrustANN}

\begin{document}
\title{Failure-Resilient and Verifiable Distributed Approximate Nearest-Neighbor Retrieval}
\author{Abdullah Al-Mamun\\Augusta University\\aalmamun@augusta.edu}
\maketitle
\begin{abstract}
Distributed vector databases increasingly support security-sensitive
applications, yet existing systems primarily protect index integrity or
replica availability rather than the trustworthiness of retrieved
results. Consequently, a compromised replica can return a plausible but
manipulated result that remains cryptographically valid, while conventional
replication and Byzantine defenses do not directly establish whether the
retrieved results are trustworthy.
We present \textsc{TrustANN}, a middleware layer that provides query-level
trust for distributed approximate nearest-neighbor (ANN) retrieval without
modifying the underlying ANN algorithm. TrustANN introduces three
complementary protocols: \textsc{TrustBind}, which verifies retrieval trust
from authenticated, consistent, and failure-domain-independent replica
evidence; \textsc{CoordShift}, which asynchronously constructs reusable
coordination epochs outside the query critical path; and
\textsc{AvailGuard}, which performs bounded recovery when replicas are
unavailable, stale, or conflicting while preserving the trust criterion.
Queries are marked \textsc{Trusted} only when sufficient independent
evidence exists for every relevant shard; otherwise, TrustANN returns
\textsc{Unverified}.
On an AWS deployment using distributed FAISS HNSW, TrustANN reduces
retrieval attack success rate from 1.000 to 0.000 under the tested
Byzantine-retrieval attack with two-of-three replica verification. The
verification layer increases clean p95 latency by only 9.36\%. Under
replica crashes, TrustANN maintains trusted retrieval through 33\% failures
and correctly returns \textsc{Unverified} when sufficient evidence is
unavailable.

\end{abstract}


\section{Introduction}
\label{sec:intro}

Distributed approximate nearest-neighbor (ANN) retrieval has become a
core substrate for retrieval-augmented generation, recommendation,
semantic search, and other vector-intensive applications
~\cite{lewis2020rag,sigmodrecord2025}. Modern ANN systems scale by
partitioning vector collections across machines and replicating shards,
while indexes such as HNSW, IVF/PQ, and DiskANN provide efficient
large-scale search~\cite{malkov2020hnsw,jegou2011pq,
subramanya2019neurips-diskann}. Recent distributed vector systems
further improve scalability through parallel execution, communication
optimization, and resource disaggregation~\cite{cotra2025,harmony2025,
distributedann2025,distvs2026}. However, distributing retrieval also
expands its trust boundary: a query result may depend on multiple
shards, replicas, and workers whose behavior cannot necessarily be
assumed correct. Thus, scalable distributed retrieval raises a
fundamental question beyond performance: \emph{when can a distributed
ANN result itself be trusted?}

First, authenticated state does not imply trustworthy retrieval.
Authenticated data structures can establish that stored state is
consistent with an authenticated representation~\cite{goodrich2001authenticated},
and recent verifiable ANN systems extend this principle to approximate
search. ANNProof, VLAH, and CD-ANN provide mechanisms for verifying ANN
search results or protecting index/vector state using authenticated
structures and, in some cases, blockchain anchoring~\cite{annproof2024,
vlah2025,cdann2026}. These approaches establish important foundations
for verifiable retrieval, but their primary abstraction is verification
of an index or an individual search execution. In a distributed ANN
fabric, however, a worker may possess a valid authenticated index while
still omitting candidates, returning stale results, manipulating
rankings, or providing inconsistent responses. Thus,
\emph{authenticated state does not necessarily imply trustworthy
distributed retrieval}. What is missing is \emph{query-level retrieval
trust}: whether the distributed participants involved in a query
provide sufficient independent evidence to justify accepting its
result. This distinction is also broader than retrieval poisoning.
PoisonedRAG and related work demonstrate that malicious retrieved
content can substantially manipulate RAG systems
~\cite{zou2025poisonedrag,raguard2026}, but a retrieval participant can
also compromise the result without injecting poisoned content, for
example by suppressing, replaying, or selectively manipulating results.

Establishing such trust, however, creates a second challenge:
coordination latency.
A natural response to untrusted participants is to require them to agree
before accepting a result. Byzantine-fault-tolerant consensus provides
strong guarantees against faulty replicas, but requires communication
and synchronization among participants~\cite{castro1999pbft}.
Blockchain-based distributed systems similarly use replicated agreement
to establish trusted state~\cite{omniledger2018,rapidchain2018}.
Putting such coordination directly on the ANN query path, however,
couples retrieval latency to distributed agreement. This conflicts with
the design objective of ANN systems, where parallelism and communication
efficiency are central to low-latency search
~\cite{malkov2020hnsw,subramanya2019neurips-diskann,cotra2025}.
Existing blockchain-anchored verifiable ANN systems demonstrate the value
of authenticated state~\cite{annproof2024,cdann2026}, but do not address
how to maintain reusable trust state without making distributed
coordination a prerequisite for each retrieval operation. The challenge,
therefore, is not simply to establish trust, but to establish
\emph{reusable trust state without placing consensus-scale coordination
on the latency-critical query path}.

Finally, failures make replica availability an insufficient
basis for trust.
Even when coordination is kept off the query path, replication provides
availability rather than sufficient trustworthy evidence. Distributed
vector systems replicate data to tolerate failures and sustain
retrieval~\cite{cotra2025,distributedann2025,distvs2026}, yet surviving
replicas may crash, become unreachable, return stale state, or produce
conflicting responses. A system that simply accepts any available
response can therefore mistake availability for correctness, whereas
lowering the integrity threshold when replicas fail weakens the very
guarantee the system is intended to provide. Byzantine-resilient systems
traditionally make the fault bound explicit when determining whether
sufficient replicas support an operation~\cite{castro1999pbft}. For
distributed ANN, the analogous requirement is to distinguish
\emph{having an answer} from \emph{having enough evidence to trust that
answer}. A trustworthy retrieval system must therefore preserve its
evidence threshold under failures, attempt bounded recovery using
surviving replicas, and explicitly reject a result when sufficient
evidence cannot be established.

\begin{table*}[!t]
\centering
\caption{Comparison of TrustANN with representative distributed,
verifiable, and Byzantine-tolerant retrieval approaches.
The listed systems address complementary system abstractions and are
not directly comparable experimental baselines for TrustANN.}
\footnotesize
\label{tab:comparison}
\begin{tabular}{lcccccc}
\toprule
\textbf{System} &
\textbf{Distributed} &
\textbf{Search/Index} &
\textbf{Replica-Level} &
\textbf{Failure-Aware} &
\textbf{Off-Path} &
\textbf{Consensus-Free} \\
&
\textbf{ANN} &
\textbf{Verification} &
\textbf{Retrieval Trust} &
\textbf{Trust} &
\textbf{Coordination} &
\textbf{Query Path} \\
\midrule
DistributedANN~\cite{distributedann2025}
    & $\checkmark$ & $\times$ & $\times$ & $\times$ & $\times$ & $\checkmark$ \\

ANNProof~\cite{annproof2024}
    & $\times$/\checkmark & $\checkmark$ & $\times$ & $\times$ & $\times$ & $\checkmark$ \\

VLAH~\cite{vlah2025}
    & $\checkmark$ & $\checkmark$ & $\times$ & $\times$ & $\times$ & $\checkmark$ \\

CD-ANN~\cite{cdann2026}
    & $\checkmark$ & $\checkmark$ & $\times$ & $\times$ & $\times$/Partial & $\times$/Partial \\

PBFT~\cite{castro1999pbft}
    & $\times$ & $\times$ & $\times$ & $\checkmark$ & $\checkmark$ & $\times$ \\

\textbf{TrustANN (This work)}
    & $\checkmark$ & $\checkmark$ & $\checkmark$ & $\checkmark$
    & $\checkmark$ & $\checkmark$ \\
\bottomrule
\end{tabular}

\vspace{0mm}
\begin{flushleft}
\footnotesize
\textit{Search/Index Verification} covers verification of search execution,
index structure, or associated state; \textit{Replica-Level Retrieval Trust}
denotes query-level trust from evidence across replicas.
\textit{Failure-Aware Trust} covers trust under replica failure, stale state,
or conflicting evidence. \textit{Off-Path Coordination} and
\textit{Consensus-Free Query Path} denote asynchronous coordination and
retrieval without consensus/global synchronization, respectively.

\end{flushleft}\vspace{0mm}
\end{table*}

\noindent\textbf{Proposed Solution.}
Existing systems address distributed ANN execution~\cite{cotra2025,harmony2025,
distributedann2025,distvs2026}, authenticated ANN state~\cite{annproof2024,
vlah2025,cdann2026}, federated retrieval confidentiality~\cite{frag2024},
or malicious retrieval content~\cite{zou2025poisonedrag,raguard2026}, but
do not provide query-level retrieval trust with reusable coordination and
failure-aware evidence recovery. We present \sys{}, a middleware layer
that operates above existing ANN indexes without modifying the underlying
search algorithm. TrustANN separates \emph{retrieval execution from
retrieval trust} and \emph{background coordination from query-level
safety} by three tightly coupled components.

First, \emph{TrustBind} establishes trust per shard from authenticated replica
evidence by requiring sufficient, mutually consistent, and
failure-domain-independent responses. Second, \emph{CoordShift} asynchronously
constructs reusable retrieval epochs through randomized,
failure-domain-aware coordination; its state is cached and consumed by
queries, but it is not the query-level safety boundary.
Finally, \emph{AvailGuard} performs bounded recovery by seeking additional
independent evidence without weakening the trust threshold. If
sufficient evidence cannot be established, TrustANN returns
\textsc{Unverified}. Thus, coordination and recovery remain supporting
mechanisms, while the query-level trust decision is independently
determined by retrieval evidence.

We implement TrustANN over an existing HNSW-based distributed ANN
substrate and evaluate it on Deep10M~\cite{zoom} using a multi-node AWS
deployment, covering retrieval attacks, evidence ablations, failures,
fan-out, recovery, and asynchronous coordination.

In summary, this paper makes the following contributions:

\noindent\textbf{(i) Query-level retrieval trust.}
We introduce TrustBind, which establishes per-shard retrieval trust from
authenticated, consistent, and failure-domain-independent replica evidence.

\noindent\textbf{ (ii) Asynchronous coordination.}
We introduce CoordShift, which builds reusable, failure-domain-aware
retrieval epochs asynchronously without placing coordination on the
query critical path.

\noindent\textbf{(iii) Bounded failure recovery.}
We introduce AvailGuard, which recovers additional independent evidence
under replica failures without weakening the trust criterion, returning
\textsc{Unverified} when sufficient evidence cannot be established.

\noindent\textbf{(iv) Prototype and empirical validation.}
We implement TrustANN over distributed HNSW and evaluate it on AWS.
The configured majority threshold reduces R-ASR from 1.000 to 0.000
against the tested Byzantine-retrieval attack, while clean p95 overhead
is 9.36\%. TrustANN maintains trusted retrieval through 33\% replica
crashes and returns \textsc{Unverified} when sufficient evidence is no
longer available.

\section{Threat Model}
\label{sec:model}

We consider a distributed ANN deployment with $N$ workers organized into
$M$ shards, each replicated across $r$ workers. TrustANN operates under
partial synchrony and evaluates retrieval trust independently for each
shard reached by a query.

\noindent\textbf{Adversary and failure model.}
We consider a Byzantine-capable adversary $\mathcal{A}$ that may control
a bounded subset of workers. For every shard $j$, fewer than half of its
$r$ replicas may be corrupted:
\begin{equation}
f_s\leq\left\lceil r/2\right\rceil-1.
\label{eq:byzantine-bound}
\end{equation}
Corrupted replicas may collude across shards and may return arbitrary,
incomplete, stale, or conflicting candidates, manipulate rankings,
replay invalid-epoch responses, or selectively respond to queries.
We separately consider crash and network failures, including unavailable,
slow, stale, or temporarily disconnected workers. Crash and Byzantine
failures may coexist within a shard. Such failures reduce available
evidence but never cause TrustANN to weaken its acceptance criteria.

\noindent\textbf{Evidence boundary.}
TrustANN evaluates evidence \emph{per shard}. For each shard $j$ reached
by query $q$, the selected evidence $S_{q,j}$ must satisfy
\begin{equation}
|S_{q,j}|\geq\theta,\qquad
C(S_{q,j})\geq\gamma,\qquad
D(S_{q,j})\geq d_{\min},
\label{eq:evidence-boundary}
\end{equation}
where $\theta$ is the support threshold, $C(\cdot)$ measures response
consistency, and $D(\cdot)$ measures independence across configured
failure domains. Evidence from different shards cannot be pooled to
satisfy these conditions. Thus, coordinated corrupted replicas across
multiple shards cannot accumulate sub-threshold evidence to satisfy a
per-shard trust decision. If sufficient evidence is unavailable,
TrustANN returns \textsc{Unverified} rather than weakening the criteria.
TrustANN does not assume that distinct worker identities imply independent
evidence; independence is determined using configured failure-domain
metadata.

\noindent\textbf{Coordination boundary.}
CoordShift provides an authenticated, asynchronously constructed
coordination reference rather than Byzantine-agreement-level
certification. A coordination round may fail or be influenced by faulty
workers without directly causing a query to be trusted: TrustBind
independently evaluates query-level retrieval evidence. A failed round
leaves the previously published epoch active and does not block queries.

\noindent\textbf{Assumptions and scope.}
We assume secure cryptographic primitives, authenticated worker identities,
and an uncompromised underlying ANN implementation. Compromise of these
mechanisms is outside our threat model. TrustANN's guarantees are bounded
by the stated per-shard corruption and failure-domain assumptions. It
does not provide a formal ANN-recall guarantee when these assumptions are
violated.

\begin{figure*}[!t]
\centering
\includegraphics[width=175mm]{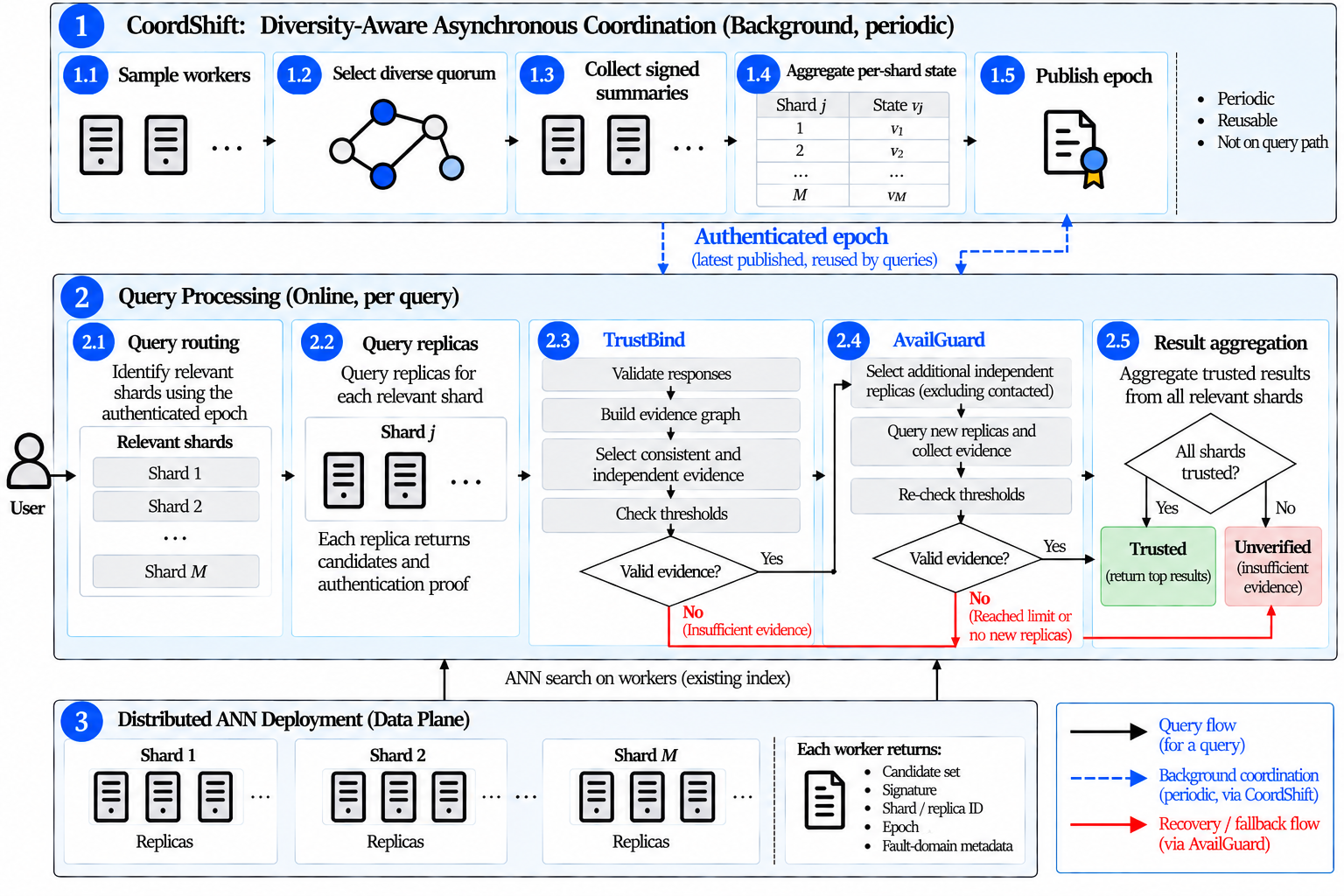}
\caption{Failure-Resilient and Verifiable Distributed Approximate
Nearest-Neighbor Retrieval.}
\label{fig:trustann}
\end{figure*}

\section{System Design}
\label{sec:design}\vspace{0mm}

TrustANN is a middleware architecture for trustworthy distributed
approximate nearest-neighbor (ANN) retrieval. Rather than replacing the
underlying ANN index or search algorithm, TrustANN operates above an
existing distributed ANN deployment and introduces a query-level
retrieval-trust layer across its shards and replicas. The design targets
environments in which retrieval workers may be unavailable, stale, or
adversarial.

The design addresses three challenges. First, authenticating ANN index
state does not establish trustworthy retrieval: a worker may possess valid
index state while returning incomplete, stale, or manipulated candidates.
Second, placing distributed agreement directly on every query introduces
additional communication and synchronization latency. Third, replica
failures can leave a query operationally available while providing
insufficient independent evidence to establish retrieval trust.

TrustANN addresses these challenges through three complementary
components: \emph{TrustBind}, \emph{CoordShift}, and \emph{AvailGuard}.
TrustBind establishes query-level trust from mutually consistent and
independently attributable replica evidence. CoordShift maintains
reusable retrieval epochs using asynchronous, randomized coordination
with fault-domain diversity, while keeping its correctness outside the
query-time safety boundary. AvailGuard performs bounded recovery by
seeking additional independent evidence without weakening the trust
threshold. Figure~\ref{fig:trustann} presents the overall architecture. \vspace{0mm}

\subsection{System Model}
\label{sec:system-model}\vspace{0mm}

We consider a distributed ANN deployment consisting of $N$ retrieval
workers organized into $M$ logical shards, with each shard replicated
across $r$ workers ($N=M\cdot r$). Each replica maintains an independent
ANN index for its assigned shard. TrustANN does not require a particular
ANN implementation; the underlying index may use HNSW, IVF, DiskANN,
or another ANN structure.

Let the distributed ANN state at epoch $t$ be
\begin{equation}
\mathcal{I}_t={I_1^t,\ldots,I_M^t},
\label{eq:ann-state}
\end{equation}
where $I_j^t$ denotes the state of shard $j$. A query $q$ is routed to a
set of relevant shards $\mathcal{J}_q\subseteq{1,\ldots,M}$ and to
one or more replicas of each relevant shard. Each contacted worker
performs ordinary ANN search and returns a candidate set together with
compact authenticated metadata identifying the worker, shard, replica,
index state, and retrieval epoch.

TrustANN operates under a partially synchronous network. Workers may
experience crash-stop failures for availability analysis and may behave
Byzantinely for integrity analysis. The adversary is bounded separately
within each shard: at most
\begin{equation}
f_s=\left\lceil\frac{r}{2}\right\rceil-1
\label{eq:byzantine-bound}
\end{equation}
of the $r$ replicas of a shard may be corrupted at a given time.
Corruption may occur independently across shards subject to this bound.
A Byzantine worker may return stale candidates, omit legitimate
candidates, inject arbitrary candidates, manipulate rankings, or provide
inconsistent evidence.

TrustANN does not require a global honest-majority assumption across all
$N$ workers. Instead, its query-level trust decision is made separately
for each relevant shard using authenticated provenance, response
consistency, and configured failure-domain independence. CoordShift
maintains reusable coordination state, but TrustBind and AvailGuard
independently determine whether sufficient retrieval evidence exists. \vspace{0mm}

\subsection{Architecture Overview}
\label{sec:architecture-overview}\vspace{0mm}

As Figure~\ref{fig:trustann} illustrates, TrustANN separates background coordination from query-level retrieval
trust. CoordShift periodically samples workers and publishes a
cryptographically authenticated retrieval epoch reference using a
diversity-aware quorum. Authentication establishes the provenance and
integrity of the epoch reference; it does not certify the semantic
correctness of its state digest. TrustBind and AvailGuard operate on the
query path, where the router selects relevant shards and replicas and
workers perform ordinary ANN retrieval.

When a query arrives, contacted workers return candidate sets and
authenticated metadata. TrustBind validates the responses and constructs
an evidence graph for each relevant shard. For each shard, it seeks a
sufficiently large set of mutually consistent responses whose provenance
spans the configured independent failure domains. A result is considered
\textsc{Trusted} only when the trust condition is satisfied for every
relevant shard.

If sufficient evidence is unavailable for one or more shards, AvailGuard
selects additional replicas for the affected shard subject to the same
independence requirements and a bounded recovery budget. It does not
lower the trust threshold and does not allow evidence from one shard to
substitute for missing evidence from another. If the required evidence
cannot be established within the recovery budget, TrustANN returns
\textsc{Unverified}.

The retrieval and coordination paths interact only through the
authenticated epoch reference, but their safety strengths are
deliberately different. CoordShift does not provide a Byzantine-safe
certification of the epoch contents. A sampled coordination quorum may
occasionally be influenced by faulty workers; such an event affects
coordination state but does not itself establish query trust. TrustBind
independently validates retrieval evidence for every query. Consequently,
a failed or adversary-influenced coordination round does not directly
cause a query to be trusted. \vspace{0mm}

\subsection{Consistency-Aware Query-Level Trust}
\label{sec:trustbind}\vspace{0mm}

\noindent\textbf{TrustBind protocol.}
An authenticated ANN index establishes the authenticity of a worker's
state but does not establish that its returned neighbors are trustworthy.
Moreover, simple replica counting can incorrectly treat mutually
inconsistent responses as equivalent evidence. TrustBind (i.e., Protocol~\ref{alg:trustbind})
therefore evaluates retrieval evidence along three dimensions:
\emph{support}, \emph{consistency}, and \emph{independence}.

\noindent\textbf{Per-shard evidence boundary.}
Because the Byzantine bound is defined per shard, the TrustBind protocol
maintains a separate evidence set for every relevant shard. Let
$S_{q,j}$ denote the evidence selected for query $q$ from shard $j$,
where $j\in\mathcal{J}_q$. Evidence from another shard cannot be used to
satisfy the threshold for shard $j$. This boundary prevents an adversary
from accumulating mutually consistent responses across different shards
and using their aggregate count to compensate for insufficient evidence
within one shard.

For each shard $j$, TrustBind accepts the retrieval evidence only if
$|S_{q,j}|\geq\theta$, $C(S_{q,j})\geq\gamma$, and
$D(S_{q,j})\geq d_{\min}$, where $\theta$ is the minimum number of
supporting replicas, $C(\cdot)$ measures response consistency, and
$D(\cdot)$ measures coverage of independent configured failure domains.
We use
$\theta=\left\lceil r/2\right\rceil$. Thus, under
Eq.~\ref{eq:byzantine-bound}, a corrupted minority within a shard cannot
alone satisfy the support threshold.

\noindent\textbf{Evidence graph.}
For each relevant shard $j$, the TrustBind protocol constructs an
evidence graph $G_{q,j}=(V_{q,j},E_{q,j})$, where each vertex represents
a valid worker response for shard $j$. An edge $(v_a,v_b)$ captures
pairwise consistency, defined as
\begin{equation}
w_{ab}=\lambda J(C_a,C_b)+(1-\lambda)R(C_a,C_b),
\label{eq:pair-consistency}
\end{equation}
where $J(\cdot,\cdot)$ is candidate-ID Jaccard similarity~\cite{jaccard}
and $R(\cdot,\cdot)$ is our normalized rank-agreement score computed over
common candidate IDs. Thus, candidate suppression primarily affects the
candidate-overlap term, whereas ranking manipulation affects the
rank-agreement term. For an evidence set $S\subseteq V_{q,j}$, aggregate
consistency is
\begin{equation}
C(S)=\frac{2}{|S|(|S|-1)}
\sum_{\substack{v_a,v_b\in S\\a<b}}w_{ab}.
\label{eq:consistency}
\end{equation}

The TrustBind protocol also evaluates evidence independence across
configured failure domains. Let $\mathcal{D}_{q,j}$ denote the failure
domains associated with replicas of shard $j$ (e.g., availability zones,
network domains, or independently managed replica groups), as specified
by trusted deployment metadata. We define
\begin{equation}
D(S)=
\left|\left\{d:\exists v\in S,\ 
\mathrm{domain}(v)=d\right\}\right|.
\label{eq:evidence-diversity}
\end{equation}
Hence, multiple responses from the same failure domain do not
automatically constitute independent evidence.

The trusted-evidence condition for shard $j$ is
\begin{equation}
\boxed{
\operatorname{Trusted}_j(q)\iff
|S_{q,j}|\geq\theta\land
C(S_{q,j})\geq\gamma\land
D(S_{q,j})\geq d_{\min}}
\label{eq:trust-condition}
\end{equation}
and a query is trusted only if every relevant shard satisfies this
condition:
\begin{equation}
\operatorname{Trusted}(q)\iff
\bigwedge_{j\in\mathcal{J}_q}
\operatorname{Trusted}_j(q).
\label{eq:global-trust}
\end{equation}

The consistency and diversity thresholds are deployment parameters. They
do not replace the per-shard corruption bound, but prevent inconsistent
or correlated responses from being treated as interchangeable evidence.

\noindent\textbf{Evidence selection.}
Finding an exact maximum-weight consistent subset is unnecessary for the
small evidence sets used by TrustANN. We therefore use a greedy
selection procedure. For each shard, the procedure begins with the
highest-supported valid response and iteratively adds the response with
the largest marginal consistency gain while respecting the configured
failure-domain constraint. The procedure terminates when the evidence
set satisfies the support, consistency, and diversity requirements or
when no eligible response remains. Thus, a response contributes to query
trust only when it is both independently attributable and sufficiently
consistent with the supporting evidence. Replica identity alone is not
treated as a sufficient indicator of independence.

\noindent\textbf{Epoch and provenance validation.}
Each response must carry a valid signature and the currently accepted
epoch number. Provenance is represented by the tuple
$\langle\mathrm{shard},\mathrm{replica},\mathrm{worker},\mathrm{domain}\rangle$,
which prevents duplicate counting while exposing the configured
independence domain used by the trust decision. The epoch reference binds
responses to the currently accepted coordination state, but the TrustBind
protocol does not treat the semantic contents of the \textsc{CoordShift}
digest as a safety oracle. Consequently, query-level trust remains
independent of whether the sampled coordination quorum produced a
perfectly accurate state summary. \vspace{0mm}

\begin{algorithm}[t]
\caption{TrustBind: Per-Shard Retrieval Evidence Verification}
\footnotesize
\label{alg:trustbind}
\begin{algorithmic}[1]
\Require Query $q$, authenticated epoch $e_t$, relevant shards
$\mathcal{J}_q$, replica sets $\{\mathcal{R}_j\}$
\Ensure Trusted evidence or \textsc{Insufficient}

\ForAll{$j\in\mathcal{J}_q$}
\State $E_{q,j}\gets\emptyset$
\ForAll{$w_i\in\mathcal{R}_j$ contacted for $q$}
\State $C_i\gets\mathrm{ANN}(w_i,q)$
\State $\sigma_i\gets\mathrm{Prove}(C_i,\mu_i,e_t,w_i)$
\State validate $\langle C_i,\sigma_i,e_t,j,\mathrm{domain}_i\rangle$
\If{valid}
\State $E_{q,j}\gets E_{q,j}\cup\{(C_i,\mathrm{domain}_i)\}$
\EndIf
\EndFor
\State construct $G_{q,j}$ from $E_{q,j}$
\State $S_{q,j}\gets
\mathrm{GreedyEvidenceSelect}(G_{q,j},\theta,\gamma,d_{\min})$
\If{$|S_{q,j}|<\theta$
\textbf{or} $C(S_{q,j})<\gamma$
\textbf{or} $D(S_{q,j})<d_{\min}$}
\State \Return \textsc{Insufficient}
\EndIf
\EndFor
\State \Return $\{S_{q,j}\}_{j\in\mathcal{J}_q}$
\end{algorithmic}
\end{algorithm}

\subsection{Diversity-Aware Asynchronous Coordination}
\label{sec:coordshift}\vspace{0mm}

\noindent\textbf{CoordShift protocol.}
Requiring Byzantine consensus or synchronous agreement for every
retrieval epoch would introduce coordination overhead into a
latency-sensitive ANN query path. The CoordShift protocol (i.e., Protocol~\ref{alg:coordshift})
instead maintains reusable coordination state asynchronously. Its purpose
is not to provide a new Byzantine consensus protocol, but to amortize
repeated coordination cost and reduce correlated exposure through
diversity-aware randomized selection.

Most importantly, CoordShift is not part of the query-time safety
boundary. Even if its sampled quorum is influenced by Byzantine workers,
the TrustBind protocol (Section~\ref{sec:trustbind}) does not accept a
retrieval result solely because it is consistent with the resulting
coordination digest.

\noindent\textbf{Diversity-aware quorum selection.}
At the beginning of coordination round $t$, the CoordShift protocol
samples a candidate pool
\begin{equation}
P_t\subseteq\mathcal{W},\qquad |P_t|=k',
\label{eq:candidate-pool}
\end{equation}
where $k'>k$. It then selects a quorum $Q_t\subseteq P_t$ of size $k$
using the greedy diversity objective
\begin{equation}
F(Q)=
\sum_{d\in\mathcal{D}}\min\{1,|Q\cap d|\},
\label{eq:coord-diversity}
\end{equation}
where $\mathcal{D}$ represents configured fault domains, such as
availability zones, network domains, or replica groups.

The selection repeatedly adds the candidate providing the largest
marginal increase in $F(Q)$. This discourages a coordination quorum from
being concentrated in a single correlated failure domain while retaining
randomized committee selection.

Each selected worker returns a signed state summary. For each shard,
CoordShift selects a majority-supported summary value when one exists;
otherwise it retains the previous value. A coordination round that fails
to obtain usable summaries is treated as a no-op, and a subsequent round
uses a newly sampled candidate pool. CoordShift therefore produces a
cryptographically authenticated, reusable \emph{coordination reference},
rather than a Byzantine-safe state certificate. Its epoch can be reused
across many queries, while query-level retrieval safety remains the
responsibility of the TrustBind protocol (Section~\ref{sec:trustbind})
and the AvailGuard protocol (Section~\ref{sec:availguard}).

\noindent\textbf{Quorum sizing and failure-domain diversity.}
The quorum size $k$ controls coordination cost and the amount of sampled
evidence, while $F(Q)$ controls representation of distinct failure
domains. Increasing $k$ increases coordination traffic; diversity-aware
selection favors distinct configured domains before filling the remaining
quorum positions. When the failure-domain metadata accurately represents
independence, this reduces concentration within correlated domains.

The design does not claim that diversity-aware sampling eliminates
Byzantine influence or universally lowers the probability of an
adversary-influenced coordination round. Instead, its directly intended
property is increased failure-domain coverage and reduced correlated
exposure. This distinction separates the coordination-resilience
objective from the query-level safety condition enforced by the TrustBind
protocol (Section~\ref{sec:trustbind}). \vspace{0mm}

\begin{algorithm}[t]
\caption{CoordShift: Diversity-Aware Asynchronous Coordination}
\footnotesize
\label{alg:coordshift}
\begin{algorithmic}[1]
\Require Membership $\mathcal{W}$, quorum size $k$, pool size $k'$
\Ensure Authenticated epoch $e_{t+1}$ or no-op
\State $P_t\gets\mathrm{RandomSample}(\mathcal{W},k')$
\State $Q_t\gets\mathrm{GreedyDiverseSelect}(P_t,k)$
\State $R_t\gets\mathrm{CollectSignedSummaries}(Q_t,\tau)$
\If{$R_t=\emptyset$}
\State \Return \textbf{no-op}
\EndIf
\State $D_{t+1}\gets\emptyset$
\ForAll{shard $j$}
\State $v_j\gets\mathrm{MajorityValue}(R_t,j)$
\If{$v_j$ exists}
\State $D_{t+1}[j]\gets v_j$
\Else
\State $D_{t+1}[j]\gets e_t.D[j]$
\EndIf
\EndFor
\State $e_{t+1}\gets\langle t+1,D_{t+1},\tau_{t+1}\rangle$
\State publish $e_{t+1}$
\State \Return $e_{t+1}$
\end{algorithmic}
\end{algorithm}

\subsection{Independence-Aware Evidence Recovery}
\label{sec:availguard}\vspace{0mm}

\noindent\textbf{AvailGuard protocol.}
Replica failures can preserve operational retrieval availability while
reducing the evidence available to establish retrieval trust. A recovery
mechanism that simply collects additional responses may incorrectly
interpret correlated or jointly compromised replicas as independent
support. The AvailGuard protocol (i.e., Protocol~\ref{alg:availguard})
therefore treats \emph{independent evidence}, rather than response count
alone, as the recovery objective.

\noindent\textbf{Design.}
Suppose the TrustBind protocol (Section~\ref{sec:trustbind}) cannot
establish $\operatorname{Trusted}_j(q)$ for a relevant shard $j$.
AvailGuard retains the valid evidence already collected for that shard
and selects additional replicas that provide new fault-domain coverage.
Replicas from domains already represented by suspicious or conflicting
evidence are deprioritized.

Each recovery step therefore attempts to increase the independence of
the evidence set rather than merely its cardinality. Importantly,
evidence recovery is performed independently for the affected shard;
responses from other shards cannot be used to satisfy the trust condition
for shard $j$. Recovery is bounded by $K_{\max}$. The threshold $\theta$
and the consistency and diversity requirements remain unchanged
throughout recovery. After each additional retrieval round, the TrustBind
protocol (Section~\ref{sec:trustbind}) reevaluates the affected shard
using the same condition as Eq.~\ref{eq:trust-condition}: \vspace{0mm}

\begin{equation}
|S_{q,j}|\geq\theta,\qquad
C(S_{q,j})\geq\gamma,\qquad
D(S_{q,j})\geq d_{\min}.
\end{equation}

If the condition becomes satisfied, the shard is marked
\textsc{Trusted}. Otherwise, AvailGuard continues while the recovery
budget and eligible independent replicas remain. If sufficient evidence
cannot be established, the affected shard is marked
\textsc{Unverified}, and the query is not labeled globally trusted.

\noindent\textbf{Safety boundary.}
AvailGuard never lowers $\theta$, relaxes the consistency threshold, or
removes the independence requirement in order to preserve availability.
Its role is limited to finding additional evidence that is structurally
independent of the evidence already collected. Thus, recovery improves
availability only when sufficient independent replicas remain available;
it does not guarantee trusted retrieval under arbitrary failure rates.
The resulting guarantee is explicitly bounded by the configured
failure-domain model. If the adversary can compromise all available
independent domains for a shard, TrustANN cannot establish retrieval
trust and returns \textsc{Unverified} rather than weakening the trust
condition. \vspace{0mm}

\begin{algorithm}[t]
\caption{AvailGuard: Independence-Aware Evidence Recovery}
\footnotesize
\label{alg:availguard}
\begin{algorithmic}[1]
\Require Query $q$, shard $j$, evidence $E_{q,j}$,
replica set $\mathcal{R}_j$, $\theta,\gamma,d_{\min},K_{\max}$
\Ensure Trusted evidence or \textsc{Unverified}
\State $b\gets0$
\Repeat
\State $S_{q,j}\gets
\mathrm{TrustBindSelect}(E_{q,j},\theta,\gamma,d_{\min})$
\If{$S_{q,j}$ satisfies the trust condition}
\State \Return $\mathrm{Trusted}(S_{q,j})$
\EndIf
\If{$b\geq K_{\max}$}
\State \Return \textsc{Unverified}
\EndIf
\State $R'\gets
\mathrm{SelectIndependentReplicas}(\mathcal{R}_j,E_{q,j})$
\If{$R'=\emptyset$}
\State \Return \textsc{Unverified}
\EndIf
\State $E_{q,j}\gets E_{q,j}\cup\mathrm{Query}(R')$
\State $b\gets b+1$
\Until{\textbf{false}}
\end{algorithmic}
\end{algorithm}
\subsection{Correctness and Safety Properties}
\label{sec:safety} \vspace{0mm}

TrustANN deliberately separates coordination properties from query-level
retrieval safety. The three protocols described in
Sections~\ref{sec:trustbind}--\ref{sec:availguard} provide the following
system-level properties.

\noindent\textbf{(1) Coordination does not determine retrieval trust.}
The CoordShift protocol (Section~\ref{sec:coordshift}) provides an
authenticated, reusable epoch reference but does not provide a
Byzantine-safe state certificate. Query-level trust is instead established
independently by the TrustBind protocol (Section~\ref{sec:trustbind});
therefore, an adversary-influenced coordination round cannot by itself
cause a retrieval result to satisfy Eq.~\ref{eq:trust-condition}.

\noindent\textbf{(2) Trust is established independently per shard.}
The TrustBind protocol (Section~\ref{sec:trustbind}) evaluates evidence
separately for every relevant shard. Under the assumed bound in
Eq.~\ref{eq:byzantine-bound}, a corrupted minority within a shard cannot
alone satisfy the support threshold, and evidence from another shard
cannot compensate for insufficient evidence.

\noindent\textbf{(3) Evidence must satisfy consistency and independence.}
TrustANN does not equate replica identity with independent evidence.
The TrustBind protocol requires sufficient response consistency and
failure-domain diversity as defined in
Eqs.~\ref{eq:pair-consistency}--\ref{eq:evidence-diversity}. Responses
from the same correlated domain therefore do not automatically increase
the trust level.

\noindent\textbf{(4) Failure recovery never weakens trust.}
The AvailGuard protocol (Section~\ref{sec:availguard}) may obtain
additional evidence after failures or insufficient evidence, but it never
lowers $\theta$ or removes the consistency and independence requirements.
If sufficient evidence cannot be established, TrustANN returns
\textsc{Unverified}. Increasing $K_{\max}$ therefore provides additional
recovery opportunities without allowing evidence from an unrelated shard
to satisfy Eq.~\ref{eq:trust-condition}.

\noindent\textbf{(5) Recovery is bounded.}
AvailGuard performs at most $K_{\max}$ additional retrieval rounds,
providing a bounded query-side recovery cost rather than waiting
indefinitely for unavailable replicas. Recovery improves availability only
when sufficient independent replicas remain available; if all available
independent domains for a relevant shard are compromised or unavailable,
TrustANN returns \textsc{Unverified}.
TrustANN does not formally guarantee that a \textsc{Trusted} result has
the same recall as an ideal unpoisoned corpus. Its guarantee concerns the
sufficiency and consistency of the observed distributed retrieval evidence
under the stated fault model. \vspace{0mm}

\subsection{Communication and Performance Model}
\label{sec:communication}\vspace{0mm}

TrustANN communication consists of three logically separate classes:
retrieval, evidence verification, and background coordination.
Retrieval traffic includes ANN requests, candidate responses, and bounded
AvailGuard fallback rounds. Evidence processing includes authentication,
provenance validation, consistency computation, and evidence-set
selection. Coordination traffic consists of state-summary requests and
responses associated with the CoordShift candidate pool and selected
quorum.

For a query serving $F=|\mathcal{J}_q|$ relevant shards with replication
factor $r$, the initial worst-case response fan-out is
\begin{equation}
B_q\leq Fr,
\label{eq:fanout}
\end{equation}
when all replicas are contacted. With selective fan-out, the actual cost
is determined by the configured initial replica set. If a full replica
set is contacted in each recovery round, a theoretical upper bound with
$K_{\max}$ additional rounds is
\begin{equation}
B_q\leq Fr(1+K_{\max}).
\label{eq:recovery-fanout}
\end{equation}
The implementation normally uses selective recovery, so these expressions
serve as worst-case communication bounds rather than measured costs.

CoordShift is amortized across queries because its epoch is reused until
superseded. Its communication and computation are therefore measured
separately from query-path retrieval and evidence processing. This
separation enables TrustANN to evaluate three distinct costs:
(1) query-level evidence verification, (2) background coordination, and
(3) failure-induced recovery. In a distributed deployment, these costs
can be measured using actual inter-node communication rather than treating
local process execution as a proxy for distributed behavior. \vspace{-6mm}

\section{Implementation}
\label{sec:implementation} \vspace{0mm}

We implement TrustANN as a middleware layer above an existing distributed
ANN index, without modifying the underlying ANN search algorithm. The
prototype~\cite{trustann-repo} is implemented in Python and comprises approximately
1.7K lines of code, excluding configuration files and
experiment scripts. The system consists of a coordinator and distributed
ANN workers connected through gRPC. Each worker hosts a replica of one ANN
shard and independently executes the local ANN search. The coordinator
implements the query router, TrustBind, AvailGuard, and the asynchronous
CoordShift service.

\noindent\textbf{Distributed ANN substrate.}
We use HNSW as the underlying ANN index, with $M=16$,
$efConstruction=200$, and $ef=64$. The ANN layer is unchanged by TrustANN.
TrustANN intercepts the retrieval result and attaches compact authenticated
evidence identifying the worker, shard, replica, and index epoch. This
separation allows TrustANN to operate as a middleware layer over existing
ANN implementations.

\noindent\textbf{TrustBind implementation.}
For each query and shard, TrustBind constructs an evidence consistency graph
whose vertices represent valid replica responses. Edge weights combine
candidate-set overlap and ranking consistency. Rather than solving a
maximum-clique problem, we use greedy evidence selection: the initial
response is chosen from the most-supported candidate set, and subsequent
responses are added according to their marginal consistency gain while
satisfying the configured failure-domain independence constraint. A shard
is trusted only when its selected evidence satisfies the configured
support threshold, consistency threshold, and failure-domain coverage
requirements. Signature and epoch checks are performed before graph
construction.

\noindent\textbf{CoordShift implementation.}
CoordShift executes asynchronously and does not participate in the query
critical path. At each coordination epoch, a seeded randomized candidate
pool is formed, and a diversity-aware selection procedure chooses the
coordination quorum. The selection favors distinct configured failure
domains before filling the remaining quorum positions. Workers return
signed state summaries, from which CoordShift derives the most strongly
supported state reference for each shard when a majority exists. The
resulting epoch is cached and reused by subsequent queries. A failed
coordination round leaves the previously accepted epoch active.

\noindent\textbf{AvailGuard implementation.}
AvailGuard handles unavailable, stale, or conflicting replica evidence.
Instead of lowering the TrustBind threshold, it selects additional replicas
that maximize independent evidence relative to the already observed
responses. Newly obtained evidence is evaluated within its corresponding
shard and cannot be pooled with unrelated shards to satisfy another shard's
threshold. Recovery terminates after a bounded number of attempts. If
sufficient independent evidence cannot be obtained, TrustANN returns
\emph{Unverified} rather than silently weakening the trust condition.

\noindent\textbf{Fault injection and deployment.}
The prototype runs as independent processes on a multi-node AWS cluster,
with each worker deployed on a separate EC2 instance and communicating
through gRPC. We use Linux traffic control and \texttt{netem} to introduce
network delay and loss. The implementation supports crash-stop failures,
unreachable replicas, stale epoch references, corrupted retrieval results,
and corrupted coordination summaries. These mechanisms allow the evaluation
to exercise TrustBind, CoordShift, and AvailGuard under both availability
and integrity faults. \vspace{0mm}

\section{Evaluation}
\label{sec:eval}\vspace{0mm}

We evaluate TrustANN on the \textbf{Deep10M} benchmark~\cite{zoom}
using a multi-node AWS deployment. We examine: (Q1) whether TrustBind
rejects corrupted retrieval; (Q2) whether CoordShift operates off the
query critical path; (Q3) whether AvailGuard maintains trustworthy
retrieval under failures; (Q4) the overhead of verification and
distributed retrieval; and (Q5) whether failure-domain-aware coordination
improves quorum diversity. \vspace{0mm}

\subsection{Experimental Setup}
\label{sec:eval-setup}\vspace{0mm}
\noindent\textbf{Dataset and ANN configuration.}
All experiments use the Deep10M benchmark~\cite{zoom}. Deep10M contains
10M 96-dimensional float32 vectors generated from deep neural-network
features. For the multi-node AWS experiments, we use a 100K-vector subset
while retaining the original 96-dimensional representation and L2
distance. Ground-truth nearest-neighbor results are used to identify
retrieval attacks. The data are partitioned across shards and replicated
across worker nodes. Each worker maintains a local FAISS HNSW index with
$M=16$, $efConstruction=200$, and query-time $ef=64$.
We use these fixed, moderate settings to provide a stable ANN substrate
while balancing search quality, construction cost, index size, and
query latency~\cite{malkov2018efficient,choi2026chat}. HNSW is used only as the underlying ANN backend
for our evaluation. TrustANN's trust layer is independent of the
underlying ANN indexing method and can be deployed above other ANN
backends. Unless otherwise stated, the deployment uses three shards with
three replicas per shard ($r{=}3$), evidence threshold
$\theta{=}2$, fallback budget $K_{\max}{=}1$, and CoordShift quorum
size $k{=}9$. The same query set and random seeds are used across
corresponding configurations.


\noindent\textbf{Baselines.}
For quantitative comparison, we use an \emph{Unverified Distributed ANN}
baseline implemented on the same distributed HNSW substrate, dataset,
workload, and AWS deployment as TrustANN, but without retrieval-level
verification. This controlled baseline provides an apples-to-apples
comparison for measuring the cost and benefit of TrustANN's query-level
trust mechanisms. The systems summarized in Table~\ref{tab:comparison}
are not used as direct experimental baselines because they target different
system abstractions and provide different verification or coordination
semantics. In particular, DistributedANN~\cite{distributedann2025}
distributes a single logical DiskANN~\cite{diskann} graph across machines, whereas
TrustANN is an index-independent trust layer evaluated here over replicated
HNSW shards. ANNProof~\cite{annproof2024} and VLAH~\cite{vlah2025} focus on
search/index verification, CD-ANN~\cite{cdann2026} incorporates Byzantine
coordination, and PBFT~\cite{castro1999pbft} provides general Byzantine
consensus rather than distributed ANN retrieval. Directly reimplementing or integrating these systems would therefore not
provide a controlled comparison of TrustANN's retrieval-trust mechanisms.

\noindent\textbf{Testbed.}
TrustANN is deployed as a multi-node cluster on AWS EC2. A dedicated coordinator hosts the query router, TrustBind verifier, AvailGuard, and CoordShift, while independent EC2 worker instances host replicated ANN indexes. Workers communicate with the coordinator through gRPC, ensuring reported query latency includes distributed communication and coordinator-side verification. Unless otherwise noted, the default deployment consists of three shards with three replicas per shard, utilizing nine worker instances (though separate scale-out experiments expand this up to 144 workers). Fault experiments terminate, isolate, or disable selected workers to emulate crash-stop and Byzantine conditions, and experiments use 100–150 queries per configuration. \vspace{0mm}

\subsection{Trustworthy Retrieval}
\label{sec:eval-trustbind}\vspace{0mm}

We first evaluate TrustBind's query-level retrieval decision. The goal is
to determine whether a query is marked \textit{trusted} (successfully 
satisfying TrustBind's per-shard evidence conditions) only when sufficient 
replica evidence is available. We evaluate security using the 
\textit{Retrieval Attack Success Rate (R-ASR)}, defined as the fraction of 
attacked queries where a manipulated or fabricated result is incorrectly 
accepted as trusted. We examine the threshold boundary, the role of the 
evidence conditions, integrity under a single-Byzantine-survivor attack, 
cross-shard evidence isolation, and additional retrieval manipulations.

\noindent\textbf{Threshold and availability boundary.}
This experiment measures the safety--availability boundary created by the
evidence threshold. We vary $\theta$ and the set of available replicas
under controlled Byzantine and availability conditions. Pattern~A leaves
only one Byzantine replica available, while Pattern~B provides two honest
replicas with one replica unavailable. We additionally test the configured
policy under clean and minority-Byzantine conditions.

\begin{table}[t]
\centering
\scriptsize
\caption{TrustBind decision boundary under Byzantine and
replica-availability conditions ($r{=}3$, 100 queries/configuration).}
\label{tab:trustbind-boundary}
\begin{tabular}{@{}lccccc@{}}
\toprule
Configuration & Scenario & Avail/Selected & Trusted & R-ASR \\
\midrule
$\theta{=}1$ & Byzantine only & 1/1 & 1.000 & 1.000 \\
$\theta{=}2$ & Byzantine only & 1/1 & 0.000 & 0.000 \\
$\theta{=}3{=}r$ & Two honest & 2/2 & 0.000 & 0.000 \\
$\theta{=}2$ & Minority Byzantine & 3/2 & 1.000 & 0.000 \\
$\theta{=}2$ & Clean & 3/2 & 1.000 & 0.000 \\
\bottomrule
\end{tabular}
\end{table}

Table~\ref{tab:trustbind-boundary} shows that with only one Byzantine
response available, $\theta{=}1$ accepts the fabricated result for every
query, yielding R-ASR $=1.000$. Increasing the threshold to the configured
$\theta{=}2$ rejects the same attack. A stricter $\theta{=}3$ also rejects
the attack but cannot establish trust when only two honest replicas are
available. In contrast, $\theta{=}2$ successfully trusts both clean
retrieval and retrieval with a Byzantine minority, with R-ASR $=0.000$.
Thus, the experiment establishes the configured threshold's safety and
availability boundary.

\noindent\textbf{Evidence-condition ablation.}
We next measure whether simply counting replica responses is sufficient for
retrieval trust. We compare three verification policies: threshold only,
threshold plus consistency, and the complete TrustBind decision rule. The
same Byzantine retrieval attack is applied while the verifier observes
the corresponding replica evidence.

\begin{table}[t]
\centering
\scriptsize
\caption{Ablation of TrustBind evidence conditions
(100 queries/configuration).}
\label{tab:evidence-ablation}
\begin{tabular}{@{}lccccc@{}}
\toprule
Policy & Selected & $C(S)$ & $D(S)$ & Trusted & R-ASR \\
\midrule
Threshold only
& 2 & 0.722 & 1.000 & 1.000 & 1.000 \\
Threshold + consistency
& 2 & 0.722 & 1.000 & 1.000 & 1.000 \\
Full TrustBind
& 1 & 1.000 & 1.000 & 0.000 & 0.000 \\
Full TrustBind, collocated
& 1 & 1.000 & 1.000 & 0.000 & 0.000 \\
\bottomrule
\end{tabular}
\end{table}

Table~\ref{tab:evidence-ablation} shows that threshold-only verification
and threshold-plus-consistency both select two responses and accept the
attacked retrieval, producing R-ASR $=1.000$. The complete TrustBind
policy instead selects an insufficient trusted evidence set and rejects
the result. The collocated configuration is likewise rejected. Hence,
additional responses alone do not substitute for the complete evidence
requirements. This experiment is an ablation of the complete decision
rule rather than independent one-factor measurements of consistency and
independence.

\noindent\textbf{Retrieval integrity.}
We then directly measure whether TrustANN prevents an attacked retrieval
from being reported as trusted. We compare an unverified distributed-ANN
baseline, TrustANN with $\theta{=}1$, and TrustANN with the configured
$\theta{=}2$ under the same single-Byzantine-survivor attack. A clean
TrustANN configuration provides a positive control.

\begin{table}[t]
\centering
\scriptsize
\caption{Retrieval integrity under a single-Byzantine-survivor
attack ($r{=}3$, 100 queries/configuration).}
\label{tab:integrity}
\begin{tabular}{@{}lcccc@{}}
\toprule
System / Condition & Trusted & R-ASR & p50 (ms) & p95 (ms) \\
\midrule
Unverified ANN 
& 1.000 & 1.000 & 0.445 & 0.516 \\
TrustANN, $\theta{=}1$
& 1.000 & 1.000 & 0.432 & 0.480 \\
TrustANN, $\theta{=}2$
& 0.000 & 0.000 & 1.535 & 1.885 \\
TrustANN, clean
& 1.000 & 0.000 & 1.523 & 1.636 \\
\bottomrule
\end{tabular}
\end{table}

Table~\ref{tab:integrity} shows that the Unverified ANN accepts the
Byzantine survivor for every query, with R-ASR $=1.000$. TrustANN with
$\theta{=}1$ exhibits the same behavior, confirming that the threshold is
essential to the protection. With $\theta{=}2$, all attacked queries are
rejected and measured R-ASR falls to zero, while the clean configuration
remains fully trusted with R-ASR $=0.000$. The protected configuration
increases p95 latency from approximately $0.52$~ms to $1.88$~ms because
replica evidence must be collected and verified.

\noindent\textbf{Cross-shard evidence isolation.}
This experiment tests whether TrustANN's recovery mechanism can
inadvertently use evidence from an unrelated shard to satisfy the trust
requirement of an attacked shard. We apply a coordinated attack to the
affected shard and vary the fallback budget
$K_{\max}\in\{0,1,2\}$.

\begin{table}[t]
\centering
\footnotesize
\caption{Cross-shard coordinated attack
($r{=}3$, $\theta{=}2$, 100 queries/configuration).}
\label{tab:cross-shard}
\begin{tabular}{@{}rcc@{}}
\toprule
$K_{\max}$ & Trusted rate & R-ASR \\
\midrule
0 & 0.000 & 0.000 \\
1 & 0.000 & 0.000 \\
2 & 0.000 & 0.000 \\
\bottomrule
\end{tabular}
\end{table}

As shown in Table~\ref{tab:cross-shard}, the attacked retrieval is never
marked trusted and R-ASR remains zero for all fallback budgets. Increasing
$K_{\max}$ therefore does not allow responses from another shard to
substitute for missing evidence in the affected shard. This confirms that
AvailGuard remains bounded by the per-shard evidence requirement.

\noindent\textbf{Additional retrieval manipulation attacks.}
Finally, we test TrustBind against two additional retrieval manipulations:
candidate suppression and ranking manipulation. Candidate suppression
removes part of the expected nearest-neighbor evidence, whereas ranking
manipulation changes the ordering of returned candidates. Clean and attack
conditions use the same queries and deployment.

\begin{table}[t]
\centering
\scriptsize
\caption{TrustANN under additional retrieval manipulation attacks
(150 queries/configuration).}
\label{tab:retrieval-attacks}
\begin{tabular}{@{}llcccc@{}}
\toprule
Attack & Condition & p50 (ms) & p95 (ms) & Trusted & R-ASR \\
\midrule
Candidate suppression
& Clean & 1.496 & 1.601 & 0.000 & 0.000 \\
Candidate suppression
& Attack & 1.496 & 1.616 & 0.000 & 0.000 \\
Ranking manipulation
& Clean & 1.496 & 1.584 & 0.000 & 0.000 \\
Ranking manipulation
& Attack & 1.500 & 1.622 & 0.000 & 0.000 \\
\bottomrule
\end{tabular}
\end{table}

Table~\ref{tab:retrieval-attacks} shows zero measured R-ASR for both attack
types under the configured TrustBind policy. Candidate suppression changes
p95 latency from $1.601$ to $1.616$~ms, while ranking manipulation changes
it from $1.584$ to $1.622$~ms. These experiments therefore provide
additional safety tests under retrieval manipulation; the primary clean
positive-control evidence remains the integrity experiment in
Table~\ref{tab:integrity}. \vspace{0mm}

\subsection{Off-Path Coordination}
\label{sec:eval-coordshift}\vspace{0mm}

We next evaluate whether CoordShift can maintain reusable coordination
state without making query processing wait for coordination. The key
property being measured is whether epoch construction is asynchronous
rather than a synchronization barrier on the retrieval path. We vary
offered query load from 50 to 400 QPS while CoordShift constructs epochs
at a fixed 200-ms interval.

\begin{table}[t]
\centering
\scriptsize
\caption{CoordShift off-path behavior
($r{=}3$, fanout$=3$, epoch interval $=200$ ms).}
\label{tab:offpath}
\begin{tabular}{@{}rccccc@{}}
\toprule
Offered QPS & Observed queries & Query p50 &
Query p95 & Commits & Waits \\
\midrule
50 & 249 & 9.277 & 10.702 & 24 & No \\
100 & 497 & 7.491 & 9.210 & 24 & No \\
200 & 750 & 6.555 & 7.993 & 24 & No \\
400 & 758 & 6.507 & 7.765 & 24 & No \\
\bottomrule
\end{tabular}
\end{table}

Table~\ref{tab:offpath} shows 24 coordination commits in every load
configuration, consistent with the fixed epoch interval. More importantly,
no query waits for epoch construction. Query p95 remains between
$7.77$ and $10.70$~ms as offered load increases from 50 to 400 QPS.
Thus, queries continue using the previously published epoch while
CoordShift constructs the next one. The observed query-count plateau at
the highest offered load reflects the capacity of the deployed retrieval
path rather than a coordination barrier. \vspace{0mm}

\subsection{Failure-Resilient Retrieval}
\label{sec:eval-failures}\vspace{0mm}

We evaluate AvailGuard under progressively increasing replica failures.
The experiment measures whether TrustANN can preserve trustworthy
retrieval when the initially contacted replicas are unavailable. We
uniformly increase the fraction of crashed replicas and record trusted
query rate, fallback activation, and query latency. AvailGuard retains
$\theta{=}2$ throughout and attempts to obtain additional independent
evidence rather than weakening the trust criterion.

\begin{table}[t]
\centering
\scriptsize
\caption{Failure resilience under uniformly random replica crashes
($r{=}3$, $\theta{=}2$, 150 queries/configuration).}
\label{tab:replica-failure}
\begin{tabular}{@{}rcccc@{}}
\toprule
Crashed & Trusted rate & Fallback & p50 (ms) & p95 (ms) \\
\midrule
0\% & 1.000 & 0.0\% & 6.631 & 7.560 \\
10\% & 1.000 & 100.0\% & 5.715 & 7.141 \\
20\% & 1.000 & 100.0\% & 5.048 & 5.904 \\
33\% & 1.000 & 100.0\% & 3.899 & 4.788 \\
49\% & 0.000 & 100.0\% & 2.596 & 3.260 \\
\bottomrule
\end{tabular}
\end{table}

Table~\ref{tab:replica-failure} shows that the trusted-query rate remains
1.000 through the 33\% crash configuration, with fallback activated at
the nonzero failure levels. At 49\% crashed replicas, however, the
trusted rate falls to zero because the remaining deployment cannot
provide sufficient evidence. Thus, AvailGuard extends retrieval
availability by locating additional evidence while preserving the same
trust requirement. The lower latency at higher crash rates results from
fewer live workers participating in retrieval and should not be
interpreted as improved system capacity. \vspace{0mm}

\subsection{Distributed-System Cost}
\label{sec:eval-cost}

We next quantify the cost introduced by distributed evidence collection
and TrustANN's verification middleware.

\noindent\textbf{Query fan-out.}
This experiment measures how retrieval breadth affects end-to-end query
cost. We vary the number of shards contacted by each query from one to
three. Since each shard has three replicas, this increases the maximum
number of replica responses processed by the coordinator from three to
nine.

\begin{table}[t]
\centering
\footnotesize
\caption{Query fan-out sensitivity
($r{=}3$, $\theta{=}2$, 150 queries/configuration).}
\label{tab:fanout}
\begin{tabular}{@{}rccccc@{}}
\toprule
Fan-out & Max evidence & Trusted rate & p50 (ms) & p95 (ms) \\
\midrule
1 & 3 & 1.000 & 1.315 & 1.513 \\
2 & 6 & 1.000 & 3.270 & 4.030 \\
3 & 9 & 1.000 & 6.583 & 7.455 \\
\bottomrule
\end{tabular}
\end{table}

Table~\ref{tab:fanout} shows that increasing fan-out from one to three
raises p95 latency from $1.51$ to $7.46$~ms and median latency from
$1.32$ to $6.58$~ms. The maximum evidence set simultaneously grows from
three to nine responses. Trusted rate remains 1.000 in all configurations,
indicating that the additional latency is associated with distributed
retrieval and evidence processing rather than failure to establish trust.

\noindent\textbf{Clean-path overhead.}
We then isolate the end-to-end overhead of the TrustANN middleware under
a clean workload. We compare the deployed TrustANN query path with an
unverified distributed-ANN baseline using the same deployment and
workload. This experiment measures performance overhead rather than
retrieval-safety behavior.

\begin{table}[t]
\centering
\footnotesize
\caption{Clean-path performance overhead of TrustANN
(150 queries/configuration).}
\label{tab:clean-overhead}
\begin{tabular}{@{}lccc@{}}
\toprule
System & QPS & p50 (ms) & p95 (ms) \\
\midrule
Unverified ANN & 1050.23 & 0.909 & 1.006 \\
TrustANN & 966.96 & 1.023 & 1.100 \\
\bottomrule
\end{tabular}
\end{table}

Table~\ref{tab:clean-overhead} shows that TrustANN achieves 966.96 QPS
compared with 1050.23 QPS for the unverified distributed ANN baseline. Median latency
increases from $0.909$ to $1.023$~ms, while p95 increases from
$1.006$ to $1.100$~ms, corresponding to a measured 9.36\% p95 increase.
These measurements quantify the clean-path cost of adding TrustANN
verification and distributed evidence processing.

\noindent\textbf{System scaling.}
To evaluate how the system behaves beyond our baseline topology, we also analyze weak scaling as the cluster size increases. Because this requires expanding the worker topology and data partitioning beyond our default 9-worker setup, this experiment utilizes a separate scale-out deployment scaling from 18 to 144 workers (with $r=3$ and $\theta=2$) under a fixed per-worker load. We report aggregate QPS, parallel efficiency, and p95 latency.

\begin{table}[t]
\centering
\footnotesize
\caption{Weak scaling on Deep10M at fixed per-worker load
($r{=}3$, $\theta{=}2$; separate scaling deployment).}
\label{tab:scaling}
\begin{tabular}{@{}rccc@{}}
\toprule
Workers & 1 & Efficiency & p95 (ms) \\
\midrule
18 & 908.6 & 1.000 & 1.258 \\
36 & 1134.0 & 0.624 & 1.159 \\
72 & 1167.0 & 0.321 & 1.436 \\
144 & 1176.5 & 0.162 & 1.482 \\
\bottomrule
\end{tabular}
\end{table}

Table~\ref{tab:scaling} shows that aggregate throughput increases from
908.6 to 1176.5 QPS as the deployment grows from 18 to 144 workers.
However, parallel efficiency decreases from 1.000 to 0.162. The results
indicate that distributed overheads increasingly limit scaling at larger
worker counts. These measurements are reported separately and are not
combined with the three-shard, three-replica AWS configuration. \vspace{0mm}

\subsection{Recovery and Reconfiguration}
\label{sec:eval-recovery}

We evaluate whether TrustANN can continue serving queries while the
distributed state is being reconfigured. We independently inject three
events---replica failure, replica replacement, and shard migration---while
queries continue to arrive. We measure the number of queries served,
query errors, query p95 latency, and total reconfiguration time.


\begin{table}[t]
\centering
\scriptsize
\caption{Recovery and reconfiguration under continuous retrieval.
Values are measured on the current AWS deployment.}
\label{tab:recovery}
\begin{tabular}{@{}lrrrr@{}}
\toprule
Event & Queries & Q. errors & Q. p95 (ms) & Reconfig. (ms) \\
\midrule
Repl. failure
& 1411 & 0 & 4.496 & 5001.805 \\
Repl. replacement
& 1323 & 0 & 4.655 & 5002.630 \\
Shard migration
& 1400 & 0 & 4.376 & 5000.342 \\
\bottomrule
\end{tabular}
\end{table}

Table~\ref{tab:recovery} shows zero query errors for all three events.
Query p95 remains between $4.38$ and $4.65$~ms, while measured
reconfiguration takes approximately $5$~s in each case. These results
show that queries continue to complete while state convergence proceeds.
Thus, reconfiguration is not implemented as a query-time synchronization
barrier. The results do not imply instantaneous convergence; rather,
they demonstrate that convergence latency is decoupled from query-path
availability. \vspace{0mm}

\subsection{CoordShift Resilience}
\label{sec:eval-coordshift-reliability}

We finally evaluate whether CoordShift's failure-domain-aware selection
changes the structural diversity of its coordination quorum under
clustered adversaries. We vary quorum size $k$, adversarial fraction $p$,
and the placement of adversarial workers across failure domains. The
primary metric is \emph{domain coverage}, defined as the number of
distinct configured failure domains represented in the selected quorum.
We also report the fraction of rounds whose selected quorum is
adversary-influenced.

\begin{table}[!t]
\centering
\scriptsize
\caption{CoordShift quorum resilience under clustered adversaries
($N{=}60$, 1500 rounds/configuration).}
\label{tab:coordshift-reliability}
\begin{tabular}{@{}rrlrr@{}}
\toprule
$k$ & $p$ & Sampling & Adv.-influenced & Domain coverage \\
\midrule
5 & .10 & Uniform
& 0.614 & 4.353 \\
5 & .10 & Diversity-aware
& 0.666 & 5.000 \\
5 & .30 & Uniform
& 0.869 & 4.329 \\
5 & .30 & Diversity-aware
& 0.941 & 5.000 \\
\midrule
9 & .10 & Uniform
& 0.827 & 6.845 \\
9 & .10 & Diversity-aware
& 0.959 & 9.000 \\
9 & .30 & Uniform
& 0.988 & 6.862 \\
9 & .30 & Diversity-aware
& 1.000 & 9.000 \\
\midrule
15 & .10 & Uniform
& 0.951 & 9.287 \\
15 & .10 & Diversity-aware
& 1.000 & 12.000 \\
15 & .30 & Uniform
& 0.999 & 9.310 \\
15 & .30 & Diversity-aware
& 1.000 & 12.000 \\
\midrule
21 & .10 & Uniform
& 0.993 & 10.677 \\
21 & .10 & Diversity-aware
& 1.000 & 12.000 \\
21 & .30 & Uniform
& 1.000 & 10.749 \\
21 & .30 & Diversity-aware
& 1.000 & 12.000 \\
\bottomrule
\end{tabular}
\end{table}

Table~\ref{tab:coordshift-reliability} shows that diversity-aware
selection consistently increases failure-domain coverage. For
$k{=}9$ and $p{=}0.10$, coverage increases from 6.845 with uniform
sampling to 9.000 with diversity-aware selection. For $k{=}15$ and
$k{=}21$, diversity-aware selection reaches the maximum configured
coverage of 12 domains, compared with approximately 9.3--10.7 domains
under uniform sampling.
The adversary-influence results in
Table~\ref{tab:coordshift-reliability}, however, require a narrower
interpretation. Diversity-aware selection does not universally reduce
the measured fraction of adversary-influenced rounds; in several
configurations, the influence rate is higher despite greater domain
coverage. Therefore, this experiment demonstrates the structural effect
of diversity-aware quorum selection on failure-domain coverage, but does
not establish a universal reduction in adversary-influenced epochs.
This coordination-layer property is distinct from query-time retrieval
safety. Even when an epoch is influenced by adversarial workers,
TrustBind independently evaluates the retrieval evidence before marking
a query trusted. CoordShift diversity and TrustBind verification
therefore serve complementary roles in the system. 
\vspace{0mm}

\section{More Related Work}
\label{sec:related} \vspace{0mm}

\noindent\textbf{Distributed ANN and verifiable retrieval.}
Distributed ANN systems such as Milvus~\cite{wang2021milvus}, CoTra~\cite{cotra2025}, HARMONY~\cite{harmony2025}, DistributedANN~\cite{distributedann2025}, DistVS~\cite{distvs2026}, and GaussDB~\cite{gaussdb2025} primarily optimize scalability, partitioning, communication, and resource management. Verifiable systems including ANNProof~\cite{annproof2024}, VLAH~\cite{vlah2025}, and CD-ANN~\cite{cdann2026} further authenticate search or index/vector state. However, these approaches generally do not determine whether \emph{distributed shards and replicas provide sufficient independent evidence to trust a retrieval result}. TrustANN instead makes the retrieval operation itself the trust boundary and establishes trust at query time.

\noindent\textbf{Federated and Byzantine-robust protection.}
C-FedRAG~\cite{addison2024cfedrag}, FRAG~\cite{zhao2024frag}, and RAGRoute~\cite{ragroute2025} address decentralized retrieval and privacy, while RAGuard~\cite{raguard2026} and PoisonedRAG~\cite{zou2025poisonedrag} target malicious retrieved content. Byzantine-robust aggregation~\cite{blanchard2017krum,yin2018byzantine,robustagg2025} and P2P defenses such as SybilGuard and X-Vine~\cite{yu2006sybilguard,mittal2012xvine,pecori2016skademlia} protect aggregation or overlay membership. These mechanisms address complementary security dimensions, but do not establish \emph{retrieval-evidence sufficiency} under replica failures, stale state, or retrieval-time manipulation. TrustANN bases its trust decision directly on authenticated replica evidence.

\noindent\textbf{Byzantine coordination and fault tolerance.}
BFT protocols and ledgers~\cite{castro1999pbft,omniledger2018,rapidchain2018,divideandscale2020} provide strong replicated agreement but introduce communication and synchronization costs unsuitable for latency-critical ANN queries. More prior work~\cite{dean,hpchain,baash,scichain,echain} similarly explored lightweight consensus, authenticated state, and provenance for trustworthy distributed computing. TrustANN instead separates coordination from retrieval: \emph{TrustBind} establishes query-level trust from replica evidence, \emph{CoordShift} maintains reusable epochs asynchronously without consensus on the query path, and \emph{AvailGuard} preserves the evidence threshold during bounded recovery. Thus, TrustANN determines whether available retrieval evidence is sufficient to trust the result without placing consensus in the ANN query path.
\vspace{0mm}

\section{Conclusion and Future Work}
\label{sec:conclusion} \vspace{0mm}

TrustANN provides a distributed trust layer for approximate nearest-neighbor
(ANN) retrieval that separates query-level retrieval trust from background
coordination and failure recovery. \textsc{TrustBind} establishes trust
from authenticated, consistent, and failure-domain-independent replica
evidence on a per-shard basis, while \textsc{CoordShift} asynchronously
maintains reusable coordination epochs without placing coordination on the
query critical path. \textsc{AvailGuard} further provides bounded,
independence-aware evidence recovery when replicas are unavailable, stale,
or conflicting, without lowering the trust threshold or relaxing the
consistency and independence requirements. As a result, TrustANN explicitly
distinguishes between a retrieval result supported by sufficient distributed
evidence and one for which such evidence cannot be established, returning
\textsc{Unverified} rather than weakening the trust decision.

Our prototype demonstrates that these mechanisms can be implemented as a
middleware layer above an existing ANN index without modifying the
underlying ANN search algorithm. Experiments on distributed FAISS HNSW
deployments show that TrustANN can reject the tested Byzantine retrieval
attack while preserving trusted retrieval under the evaluated replica
failure conditions, with a 9.36\% increase in clean p95 latency. More
broadly, the results demonstrate that retrieval-level trust can be
established through explicit evidence requirements while keeping global
coordination and recovery decoupled from the latency-critical retrieval
path.

As future work, we will investigate adaptive adversaries, larger-scale and
more dynamic failure and reconfiguration scenarios, and stronger
retrieval-level guarantees. These directions will extend TrustANN's
evidence-based trust model to increasingly dynamic and adversarial
distributed retrieval environments.



\bibliographystyle{plain}
\bibliography{paper_refs}
\balance
\end{document}